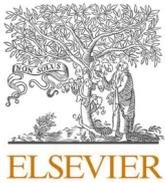
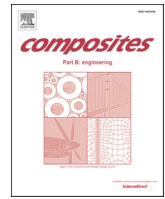

# Directional eddy current probe configuration for in-line detection of out-of-plane wrinkles

Meirbek Mussatayev [a,*], Qiuji Yi [b], Mark Fitzgerald [c], Vincent K. Maes [d], Paul Wilcox [a], Robert Hughes [a]

[a] *Department of Mechanical Engineering, Ultrasonics and Non-destructive Testing (UNDT), University of Bristol, Bristol, BS8 1TR, UK*
[b] *Department of Computer and Information Sciences, Northumbria University, UK*
[c] *Electrical Workshop, Engineering Faculty Office, University of Bristol, UK*
[d] *Department of Aerospace Engineering, Bristol Composites Institute (ACCIS), University of Bristol, UK*

ABSTRACT

Real-time monitoring of carbon fibre composites during Automated Fibre Placement (AFP) manufacturing remains a challenge for non-destructive evaluation (NDE) techniques. A directional eddy-current (EC) probe with asymmetric transmit and differential receive (Tx-dRx) coils is designed, constructed and characterized to evaluate the detectability of out-of-plane wrinkles. Initial studies were conducted to determine suitable excitation frequencies and to analyse the impact of relative orientations of driver and pickup coils on wrinkle detectability. The probe configurations are evaluated experimentally and employ a new finite element modelling approach to better understand the relationship between eddy-current density and defect detection. The findings indicate that a probe configuration with an asymmetric driver coil normal to the material surface and aligned with the fibre directions, and with differential pickup coils 90° to the scanning direction, shows the best capability for out-of-plane wrinkle detection, with SNR >20 for wrinkles over 1.3 mm in amplitude.

## 1. Introduction

Carbon fibre reinforced polymer (CFRP) composite materials are made from high-strength carbon fibres embedded in the flexible and tough epoxy matrix. They are widely used and attractive for many industrial applications because of their superior mechanical properties in comparison with its conventional counterparts [1–3]. CFRP provides the best of both worlds - high strength from carbon fibres, with flexibility and toughness from the epoxy matrix. CFRP engineering components are regularly manufactured by stacking several ply laminate layers, with distinct fibre orientations, to obtain suitable mechanical properties along desired load paths. The exponential growth in demand for CFRP production over the past decade from 2010 to 2020, and predictions of further growth in global demand by 2050 [4] is clear evidence for the need to increase automated production rates.

Automated fibre placement (AFP) and automated tape layup (ATL) are the most common methods of CFRP production, offering minimal material scrap, low labor costs and high repeatability, accuracy and productivity in comparison with manual composite manufacturing methods [5]. The current challenge in AFP-based manufacturing processes is the accurate inspection and quality monitoring of the as-laid material – something currently performed by qualitative and subjective manual visual methods.

Any potential in-line inspection technique for AFP manufacturing should meet the following prerequisites:

1. Non-contact - to prevent the introduction of contact damage or foreign contaminants into the layup structure.
2. Compatibility with AFP process environments - i.e. high deposition speeds (0.1–1 m/s) [6], elevated temperatures, and variable stand-off distances from the surface.
3. Sensitive to critical flaws - including misalignment of fibers and inter-laminar flaws.
4. Safety, size and weight limitations – to allow ease of implementation and retrofitting to existing AFP technologies.

Out-of-plane wrinkling typically arises during the manufacturing process, and causes a significant degradation to mechanical performance of the composites. The research conducted by Potter et al. [7] indicates that a clear correlation exists between the reduction in strength and the amplitude of wrinkling, and can result in a reduction in the strength of up to 50 % in compression and up to 70 % in tension. Simulated studies show that all plies affected by out-of-plane wrinkling led to a reduction of 72 % in stress at initiation of damage [7]. It is common during the formation of multiple plies around a simple radius for wrinkling to occur since plies cannot slip over one another because of path length differences between the upper and lower side of the ply [8]. The origins of out-of-plane wrinkling can be caused by many other factors and their occurrence is extensively described in Ref. [8].

The realisation of in-line monitoring would allow several possible improvements of AFP manufacturing performances, namely:

- Detect defects at an early stage.
- Full automation of AFP production and improving manufacturing performance via closed-loop control.
- Build-up of an as-manufactured "digital twin" of parts to help for mechanical modelling predictions.

Many authors have approached monitoring and inspection of AFP manufactured plies. There are several initial studies towards in-situ AFP inspection and the main NDE techniques are thermography and profilometry. Some successful implementations of in-line AFP systems were recorded at the early stage by commercial companies such as Flightware [9], Coriolis [10] and Electrimpact [11] for in-line inspection of AFP systems. Juarez et al. [12] recently presented a well-developed thermography inspection technique. Highly sophisticated hardware along with a complex robotic system were exploited to create an advanced inspection system. Recent advances in AFP inspection demonstrated the feasibility of the "*observe-think-react*" approach in Ref. [13], showing the potential of the real-time automated fibre placement (RT-AFP) machine in detecting out-of-plane wrinkles and preventing it by adjusting the process parameters on-the-go. Various types and sizes of defects should be tested to further develop the proposed RT-AFP machine. A summary of recent advances in in-situ inspection approaches is given in Table 1.

Eddy current testing (ECT) techniques, are already used for the inspection of fibre-related defects and are one of the most suitable NDE techniques for real-time monitoring of a composite structure [20–22].

**Table 1**
Summary of recent advances in AFP inspection.

| Inspection technique/NDT Method | Industrial Application | Motivation | Types of defects | Limitations | Advantages | Validation | Proposed in-situ device |
|---|---|---|---|---|---|---|---|
| Profilometry [9] | 100 % inspection during continuous production at AFP speed up to 2 m/s | Decreasing the cost of AFP production | Gaps, overlaps, twisted tows and fuzzballs | The location of defects smaller than 0.5 mm | Increase productivity by 20–30 % compared to current methods | None. | A line laser and a camera |
| Thermography [14] Thermography [15] Thermography [16] Thermography [17] Thermography [12] | Inspecting areas of insufficient ply adhesion | Limitation of laser profilometry | Laps, gaps, and twists | A lower rate of capture, i.e. low speed of layup | To reduce risk and increase laminate quality during fabrication | None. | An uncooled microbolometer sensor |
| | Interruption of AFP due to manual QA (like visual inspection) | Time and cost | Gaps/overlaps, twisted and spliced tows | The computing speed of the developed algorithm limit the real-time capability of the monitoring system. | Online process monitoring system to detect defects for lay-up speeds up to 1 m/s | None. | IR camera combined with image processing |
| | Time-consuming and insufficiency of visual inspection | Time and quality | Tow position, foreign bodies, gaps/overlaps | Cooling tows prior to inspection to create temperature contrast between surface & tows | Localize position of the single lay-up courses in each ply and detect gap width | Comparison between actual and planned positions and estimation of tow and gap widths | A thermal camera |
| | Creating more sufficient production environment | Decreasing the cost of inspection | Tow overlap/gap, wrinkling, and peel-up, poor/loss of adhesion between plies and the effects of vacuum debulking | To heat the substrate (base layer) prior to lay up | Capable to detect visually undetectable defects: poor/loss of adhesion between plies, the effects of vacuum debulking | None. | Infrared camera, quartz heating lamp |
| | The inspection techniques are ex situ with certain limitations to some type of defects | Detection limitations of certain type of inspection techniques | Gaps, overlapping tows, and Foreign Object Debris | The inspection speed is related to the frame rate of the camera | Enable high-rate AFP manufacturing with minimal interference; novel calibration regime | Comparison with artificial manufacturing induced defects | In Situ Thermal Inspection System |
| ECT [18] | Automated process monitoring of AFP | Time and cost | Fiber orientations | Penetration depth and resolution. | Applicable on rough surfaces | None. | A differential probe, a half transmission probe (HT) and a high-frequency probe (HF) were tested |
| The real-time AFP machine (profilometry-based sensors) [19] | Active process control and defect detection during AFP manufacturing | Early defect detection and correction in real-time in AFP manufacturing | Out-of-plane wrinkles | The targeted defect size was 3.2 mm | The machine can automatically adjust the process parameters to correct the defects | Comparison with a microscopic image | The profilometry-based laser line scanners |

Firstly ECT techniques do not require contact with the surface. Secondly they are capable of high inspection speeds, up to 4 m/s [23,24]. Another advantage is that post-processing of ECT data is not computationally demanding and provides quantitative measurements, making the characterisation of defects related to size, shape, location and depth clearer and traceable. Furthermore, it can operate over a wide frequency range from kHz up to a 10s of MHz, where the highest sensitivity to defects within CFRP is usually expected [25]. A final advantage is that the ECT instrumentation can be made comparatively small in footprint and each device is easily replaceable which means further miniaturization is possible. As a result, ECT was chosen as a suitable NDE technique for in-line inspection.

This paper reports on the development of ECT sensors suitable for in-line, real-time quality monitoring of AFP manufacturing flaws and to explore the design parameters of the novel directional ECT probe capable of detecting out-of-plane wrinkles. Section 2 introduces the methods including sensor design, configurations and operation mode. The finite element modelling (FEM) approach is explained in Section 3, with predictions for the induced eddy-currents (ECs) for each probe configuration. The experimental study into out-of-plane wrinkle detection and validation of the FEM study is presented and discussed in Section 4. Section 5 concludes with the main findings and future potential of the research.

## 2. Methods

The concept of embedding ECT systems into the AFP head is based on a directional probe design introduced by Kosukegava et al. [26]. While conventional EC probes are very sensitive to lift-off variations, the hybrid transmit and differential receive (Tx-dRx) nature of the proposed probe enables the sensor to minimize the effect of lift-off due to the movement of the AFP head. Moreover, the unidirectional conductivity of the CFRP layers results in strong EC densities along the fibre directions of the specimen. Thus, to support the ECs to flow widely and uniformly the uniform driver coil is exploited [26]. The research in the current paper comprises a series of studies to find a suitable configuration from four anisotropic probe designs by testing them on out-of-plane waviness in cured CFRP.

### 2.1. Eddy current theory

Eddy-Current Testing (ECT) is reliant on Faradays' law of electromagnetic induction where a changing magnetic field will generate eddy currents (ECs) when incident upon an electrically conductive material. The induced ECs generate their own secondary magnetic field which carries information about the conductivity and magnetic permeability of the test material, and tries to drive its own current in any inductors at the surface. Specific information about the specimen can therefore be revealed by monitoring variations in the electrical properties of pick-up coils above the material under the test [27]. The conductivity of carbon fibre allows for the application of ECT techniques for inspecting CFRP for defects [28–30]. Fig. 1 shows one of proposed probe configurations and the expected EC flow based on ECT theory. Due to the low electrical conductivity of composites, weak detection signals are observed in traditional EC inspection. However, by performing high frequency measurements between 100 kHz and 100 MHz [31], sufficient current is induced to obtain high-resolution imaging of fibres and other features. In addition, operating probes around electrical resonance enhances the sensitivity of ECT to features within the material [32–35].

Directional eddy-current probes with Tx-dRx coils were developed to evaluate the detectability of out-of-plane wrinkling, inspired by directional probes proposed in Ref. [26], where the primary magnetic flux is distributed in the area below the probe [36]. Due to the anisotropic nature of the composites, the desired probe configuration is expected to induce ECs to flow along the fibres' orientation, and, by placing differential sensing coils one after another in the scanning direction, a strong defect response is possible while minimizing the effects of lift-off. The next section provides a detailed explanation of each component of the probe, and the configurations evaluated in the study.

### 2.2. Directional probe design

A schematic diagram of the proposed probe with its dimensions and relative positions is illustrated in Fig. 1. In this study the authors developed and tested a directional EC probe operated in Tx-dRx mode, inspired by Kosukegava et al. [26]. Unidirectional CFRP plies have high conductivity along the fibre direction but low conductivity in the transverse direction, thus, the uniform magnetic field created by the driver coil as is shown in Fig. 1 will induce current to flow along with the fibre directions (opposing the direction of coil windings). The rectangular pickup coil is designed with elongated pickup coils orientated orthogonal to the scanning direction (see Fig. 1) such that the defective zone will generate asymmetric ECs flows beneath the differential coils [37] and the voltage difference between two adjacent pickup coils is then measured as shown. The differential voltage is generated when one of the differential coils senses a non-defective area while the other one is above a defective zone. Therefore, the pickup coils are orientated orthogonal to the scanning direction (please note that the direction of fibres on the top surface of an AFP manufacturing system would be parallel to the scanning direction as is shown in Fig. 1(c)). Having the elongated length of the pickup coils parallel to the scanning direction will significantly decrease the targeted defect detection due to signal cancellation when both pickup coils sense the defective zone equally. This section provides an explanation of proposed probe design in detail.

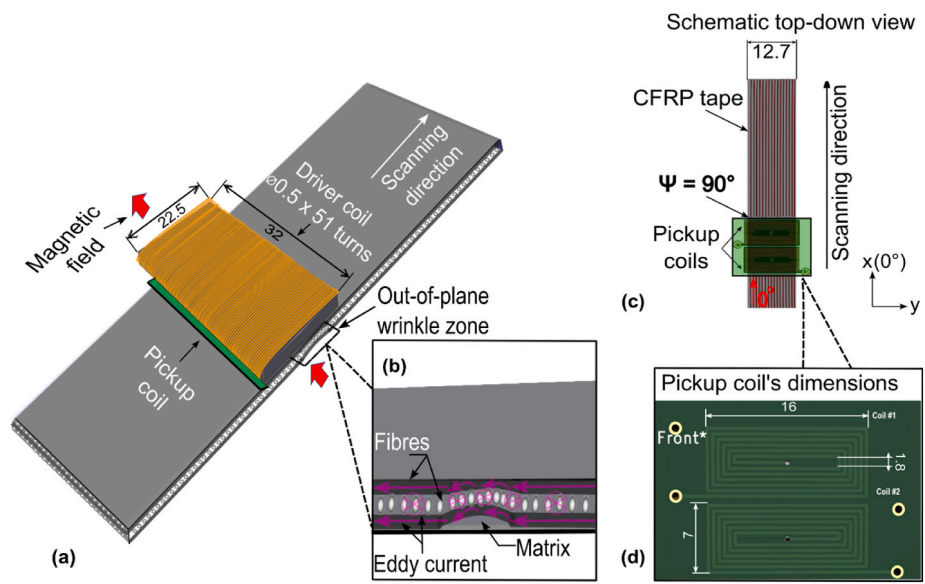

**Fig. 1.** Example configuration of directional EC probe on AFP deposited substrate with asymmetric Tx-dRx coils' dimensions: (a) Induction of ECs within a cross-ply CFRP material and dimensions of the directional EC probe design; (b) zoomed in side-view of ECs directions; (c) schematic top-down view; (d) rectangular planar pickup coils' dimensions.

*2.2.1. Driver coil*

The driver coil is hand wound around a rectangular ferrite core (Ferroxcube PLT38/25/3.8–3F4) with 51 turns of 0.5 mm diameter wire as shown in Fig. 1(a). The solid ferrite core increases the primary magnetic flux density around the pickup coils. The dimensions and properties of the driver coil are shown in Table 2. The inductance of both driver and pickup coils were measured using the inductance, capacitance, and impedance meter (PM 6303 RCL, Philips, Netherlands). The targeted inspection width of the AFP based composite is 12.7 mm. Thus, it is not necessity for the probe to cover larger areas in this application. The impedance spectra of the coils were obtained by using a Network Analyser (TE3001, Trewmac Systems, Australia).

*2.2.2. Pickup coil*

The design of the proposed pickup coil consists of two rectangular printed circuit board (PCB) coils, with $16 \times 7.5$ mm dimensions and a distance between the coils of 1 mm. Built on an FR-4 substrate of 1.6 mm height, the traces are 0.3 mm wide with a 0.2 mm pitch, yielding a total of 12 turns across 2 layers PCB as is shown in Table 2.

*2.3. Probe configurations*

Due to the unknown relationship between fibre orientation and the electrical properties it is necessary to optimize the relative orientations of driver and pickup coils to increase their sensitivity to perturbations of ECs caused by defects which lead to changes in the secondary magnetic field [28]. Thus, four probe configurations were examined with varying relative orientations of driver to the pickup coils' as is shown in Fig. 2. The notation of each probe configuration is based on the following: The $0^\circ$ and the scanning axis (direction of movement) in AFP layup is always aligned with the fibre direction. Cartesian axes (x, y, z) are defined with x and y in the plane of the composite, and with x aligned to the $0^\circ$ direction. For all probe configurations considered in this paper, the differential pickup coils are orientated at $90^\circ$ to the scanning direction i.e.

**Table 2**
Material properties and dimensions of the driver and pickup coils.

| Driver coil | Parameters and dimensions | Pickup coil | Parameters and dimensions |
| --- | --- | --- | --- |
| Number of turns | 51 | Number of turns | 12 |
| Wire diameter | 0.5 mm | Layers | 2 |
| Dimensions | $32.0 \times 22.5 \times 5.5$ mm | Copper Weight | 10.68 kg/m$^2$ |
| Inductance | 102 $\mu$H | Height | 1.6 mm |
| | | Material | FR4 (150 °C) middle Tg |
| | | Circuit Size X | 16 mm |
| | | Circuit Size Y | 7.5 mm |
| | | Inductance of each coil | 1.2 $\mu$H |

the longer dimension of the coils are perpendicular to the x-axis (see Fig. 1). The letters "uni" and "non-uni" in the probe naming convention used is denoted by the word "uniform" and "non-uniform", respectively. The first number in the naming convention, either "0" or "90", refers to the primary direction of ECs flow generated by the driver coil, relative to the scanning direction. Finally, the letter "p" and "o" refers to the parallel and orthogonal orientations of the driver coil B-field relative to the sample, respectively.

Fig. 2 shows the different probe configurations evaluated. Probes uni_90_p and uni_0_p generate uniform magnetic fields resulting in uniform EC induction parallel and orthogonal to the driver coil winding directions respectively, as is shown in Fig. 2(a) and Fig. 2(b). In probe uni_90_p (Fig. 2(a)) the driver is generating a uniform magnetic field and it is trying to drive currents in a single direction along the fibres resulting in relatively strong current densities generated in the top surface of the sample. The non-uni_90_o probe (see Fig. 2(c)) is also driving current along the fibres in the top surface but with the driver coil orthogonal to the surface of the sample. Consequently, in this scenario, the ECs flow along the fibres which is expected to generate a much higher EC density. The probes uni_0_p and non-uni_0_o (see Fig. 2(b) and (d), respectively) were designed with the driver coil generating ECs across the top surface of the material. So, in the normal position to the surface the excitation coil's non-uniform magnetic field is pointing down and it is driving the current directly beneath the pickup coils.

In the following section these probes will be evaluated on cured samples while varying the severity of targeted defect. To begin with Finite Element Modelling (FEM) of the four proposed configurations is used to virtually test their responses to discontinuities in samples, and the detectability is quantitatively analyzed in the following sections.

*2.4. Resonant frequencies of probes*

The resonant frequencies of both driver coil in air and pickup coils when assembled in the four configurations are shown in Fig. 3(a) and Fig. 3(b), respectively. The hybrid asymmetric transmit and differential receive nature of the proposed probe is different from the absolute probe of conventional ECT where operating above the resonance frequency exhibits sharp decrease in sensitivity with increasing frequency. Due to this special design, the directional EC probe can explore a broader frequency measurement range (see vertical straight lines in Fig. 3) around the resonance and can detect defective signals with a high SNR. The precise mechanism for this is not yet known, but a previous paper by Ref. [26] highlights that sensitivity is highest just beyond resonance in this configuration. Based on the findings in Ref. [26] optimum sensitivities were observed when the driver and pickup resonant frequencies were comparable. Consequently, by applying external capacitors the resonance of the pickup coils were tuned to resonate at a lower frequency as suggested in Ref. [26] of the sensing coil to make the pickup signal response as high as possible, this study aims to explore the sensitivity of different anisotropic probe configurations to out-of-plane wrinkling within the measurement range around resonance between 1.4 and 2.3 MHz. The original resonant frequency of the un-capacitively-loaded pickup coils were $13.6 \pm 0.1$ MHz in free space. The resonances of the sensing coils remain consistent due to being PCB manufactured, with only a minor difference between peak amplitudes, $|Z(f_0)| = 29 \pm 4$ Ω. The driver coil's resonance occurred at 1.59 MHz. Tuning the frequency to below the driver coil's resonant frequency still allowed the measurement of the same structural response from the sample. Thus, a 10 nF capacitor was used to capacitively load [38] the pickup coils of each probe configuration so that they resonated at 1.5 MHz as is illustrated in Fig. 3 b.

**3. FEM of EC probe configurations**

An initial finite element modelling (FEM) study was performed to understand how the orientation of the driver coil influences current densities in the top surface and between the layers (at the interface) of the CFRP sample.

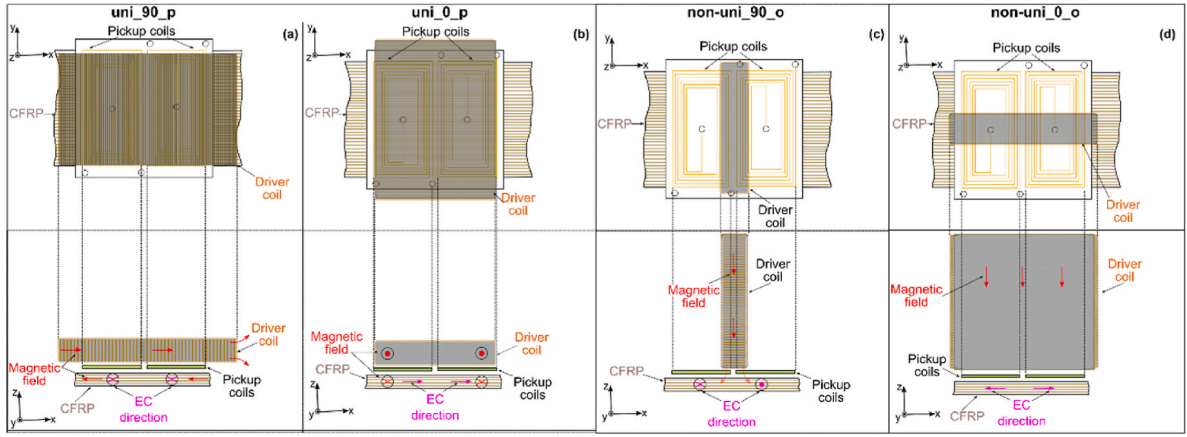

**Fig. 2.** Schematic top and side view of probe configurations (left to right): (a) uni_90_p, (b) uni_0_p, (c) non-uni_90_o and (d) non-uni_0_o probe.

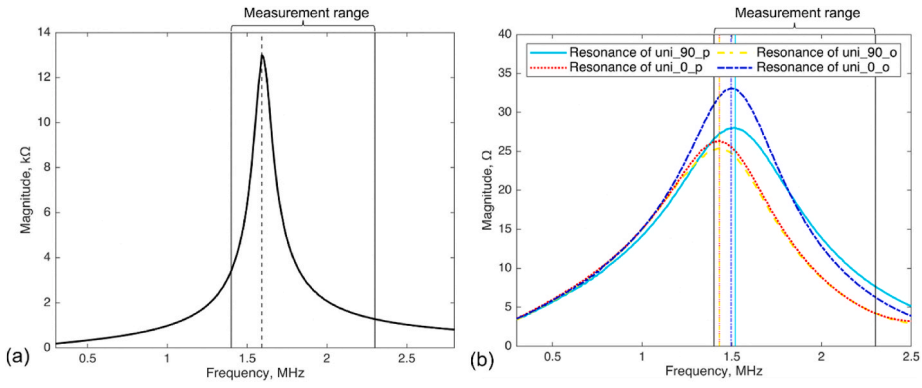

**Fig. 3.** Impedance spectra measurements for: (a) the driver coil in air; (b) pickup coils after 10 nF capacitive loading in the four probe configurations placed on undamaged CFRP.

*3.1. FEM method of low conductivity tensor of CFRP*

The CFRP structure is complex and exhibits electrical anisotropy which can affect the ability of an EC in-line monitoring system to detect defects if not considered. In this section, the modelling approach developed by Yi et al. [39] is implemented to simulate structural variation in the conductivity tensor representing coherent structural noise. A wrinkle is defined as a region of lower electrical conductivity integrated within the structural model to emulate a realistic coupon sample with a symmetric wrinkle defect. The electromagnetic numerical simulation was developed using COMSOL 5.6 and is shown in Fig. 4(a).

To investigate the electrical response generated in CFRP, the electrical conductivity tensor must be defined. It should be noted that the $\sigma_L$ is the conductivity parallel to the fibre direction, $\sigma_T$ denotes that in the transverse direction and $\sigma_{cp}$ refers to the cross-ply (through-thickness) conductivity. Therefore, for unidirectional CFRP, the homogenised conductivity tensor as a function of orientation $\theta_f$ (fibre orientation) in a given ply layer can be introduced as follows [40],

$$\sigma = \begin{Bmatrix} \sigma_L \cos^2 \theta_f + \sigma_T \sin^2\theta_f & \frac{\sigma_L - \sigma_T}{2} \sin 2\theta_f & 0 \\ \frac{\sigma_L - \sigma_T}{2} \sin 2\theta_f & \sigma_L \sin^2 \theta_f + \sigma_T \cos^2 \theta_f & 0 \\ 0 & 0 & \sigma_{cp} \end{Bmatrix}. \quad (1)$$

The tensor modulation method [39] is used to simulate the structural variation of the CFRP laminates. The detailed implementation of the CFRP modelling can be found in Ref. [39]. Therefore, by combining the Tx-dRx setup and spatially modulated conductivity tensor the FEM is built. To ensure fast convergence, the coil geometry analysis is first conducted with a relative tolerance of 0.001 and the analysis of the frequency domain is conducted in the second stage. Fig. 4(b) demonstrates the model mesh. The numeric multi-turn homogenised coil model strategy is implemented to simulate the rectangular driving coil, as shown in Fig. 4(c), while the arrow denotes the current flow direction.

Table 3 displays the main parameters used in the FEM modeling. The whole model has 2,277,196 DOF. The element size was reduced until convergence was achieved. The size of the element ranges from 0.5 mm to 4 mm, depending on position in the model. The whole model is computed on a PC with Ryzen 5900 with 12 cores.

*3.2. FEM probe configuration analysis*

Virtual testing of simulated wrinkles integrated with structural variation of the conductivity tensor [39] is performed. This section quantitatively analyses the performance of each probe configuration using this simulated data.

FEM was used to simulate the four different Tx-dRx configurations (see Fig. 5(a)-5.(d)) above a bi-directional CFRP 2-layer specimen with simulated out-of-plane wrinkling (10 mm wavelength), simulated as a reduction in the conductivity of 0 % (defect free with modulation) 25 %, 50 % and 75 % compared to unmodulated material with 50 % drop of conductivity. The FEM results shown in Fig. 5 are the differential voltage between pick-up coil 1 (Rx1) and 2 (Rx2). The previous work mentioned that the effectiveness of the modelling is only validated on the resistive component, therefore simulated data are all real parts of the voltage [39]. The SNR is used to compare the simulated performance of these configurations based on the peak-to-peak voltage relative to the RMS value of the background structural noise (N$_{rms}$) [34], as,

$$SNR = \frac{V_{pp}}{2N_{rms}}, \quad (2)$$

where $V_{pp}$ is the defective voltage (see Fig. 5(b)). It is noted that all four probe configurations except uni_0_p clearly demonstrate the expected symmetric pattern combined with texture/structure noise and a wrinkle defect response (Fig. 5). It is noted that the simulated fibre tow response (structural noise) is observable in the undamaged area. Therefore, it is feasible to quantify the SNR using the RMS of the undamaged area and the wrinkle response from the central conductivity modulated region.

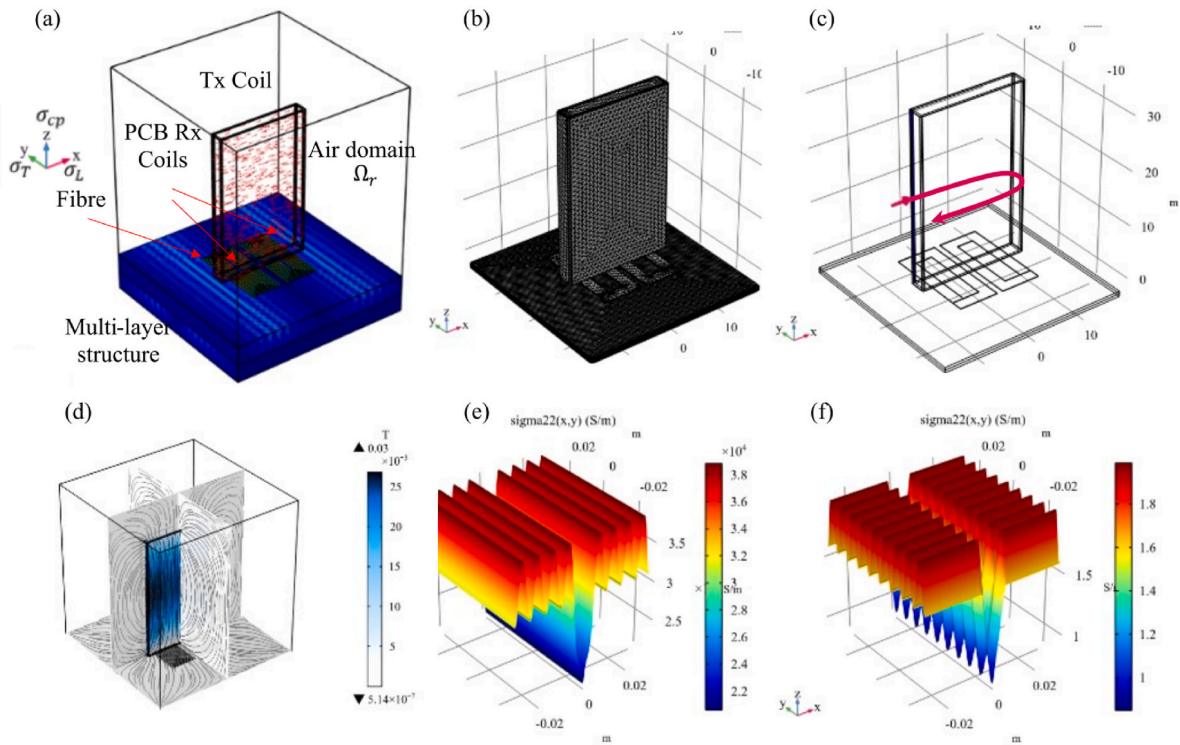

**Fig. 4.** FEM modelling for Tx-dRx ECT system: (a) example solution model, (b) model mesh, (c) numeric analysis of Tx coil geometry, (d) magnetic flux distribution of Tx coil, (g) $\sigma_{22}$ of 90° layer, (h) $\sigma_{11}$ of 0° layer.

**Table 3**
Parameters used for FEM modelling.

| Parameters | Values |
| --- | --- |
| Driving coil | 51 turns, 31.75 mm × 20.32 mm ×3.18 mm |
| Rx coil | 12 turns (over two layers), 0.3 mm width and 0.2 mm spacing |
| Core material | BMHFi-6/23 B |
| Lift-Off | Rx to sample: 0.1 mm, Tx-Rx: 0.5 mm |
| $\sigma_L, \sigma_T$ | 39000 $S/m$, 7.9 $S/m$ [41], fibre tow wavelength 6.6 mm [39] |
| Sample thickness | [90/0]$_2$, each layer is 0.5 mm thick. |
| Wrinkle properties | $\lambda = 10$ mm, positioned at each layer |
| | The wrinkle of different severity is simulated as %drop (0 %–75 %) of $\sigma$ at this area |
| Frequency | 1.9 MHz |
| COMSOL module | AC/DC module |
| Physics setting in COMSOL | Magnetic fields |
| | Electric currents in shells |
| | Homogenised multi-turn for coil |
| FEM study steps | Study1: Coil Geometry analysis |
| | Study 2: Frequency Domain study |

The uni_90_p probe appears to exhibit the worst performance for defect detection. Due to the symmetry of the differential pickup coils, the peaks measured by all orientations of the driver coil are symmetric, as shown in Fig. 5(a). Parallel configurations induce higher currents on the top surface, while orthogonal probe configurations have higher voltage values on the interface. The voltage values also indicate that a smaller footprint for inducing currents leads to an excellent capability for detecting local wrinkles.

The current densities were analyzed to evaluate the interaction between coils and the sample for the different probe configurations. The current density distribution of the top surface and at the interface (0.25 mm depth) were simulated. Firstly, the wrinkle area with lower conductivity is observed at the interface. The highest contrast of current density can be found in the first row of Fig. 5(c), where the dominant current axis flows in the direction of fibres and Rx coils are orthogonally aligned. Secondly, exciting samples with the larger area and uniform B-field and the higher current density, as demonstrated in Fig. 5(b) which has a current density two times larger. Thirdly, when current flows in the 0° (fibre direction of the second layer), the interface current density is higher than that of the top layer for probes non-uni_0_o and uni_0_p. This effect could be used to excite a stronger electrical response for deep layers in a realistic application. The data collected from both simulations and experimental studies are not expected to be directly equivalent due to the complexity of the material and defects.

Fig. 5 shows: i) FEM representation of the four different Tx-dRx probe configurations and ii) finite element scan results of simulated 1D wrinkle defects on [90/0] CFRP structure with varying conductivity regions simulating a wrinkle region. Exploring the 4 probe configurations in FEM, allows for the virtual assessment of defect detectability for a given probe configuration. To validate the FEM findings, experimental studies were performed on manufactured out-of-plane wrinkles.

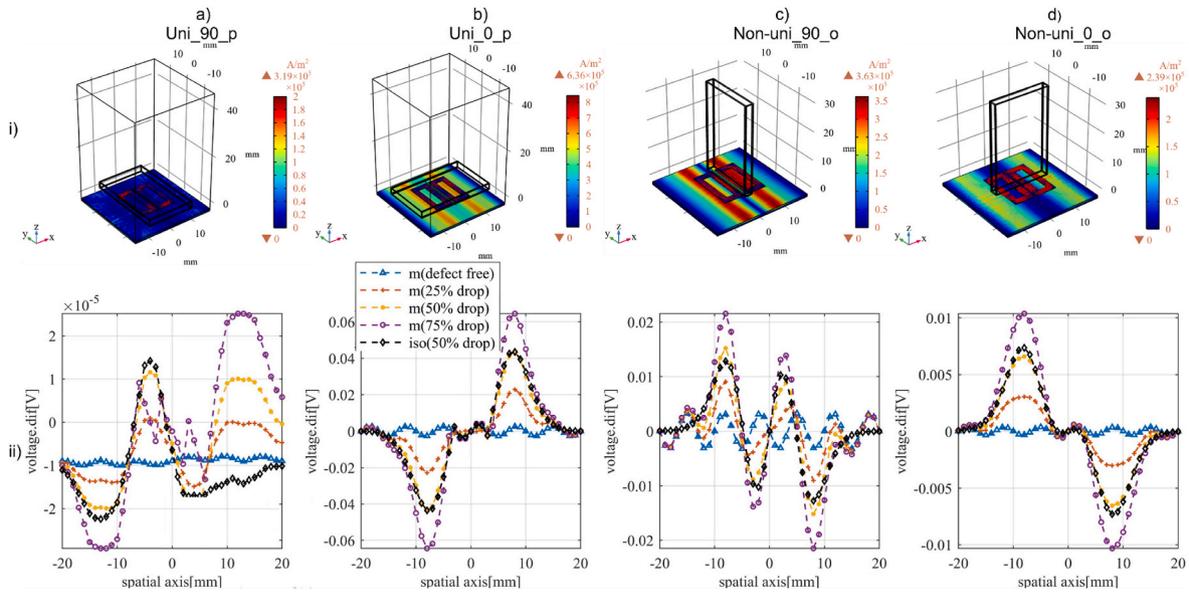

**Fig. 5.** FEM study results for evaluating the best Tx-dRx ECT sensor configuration using: (a) uni_90_p; (b) uni_0_p; (c) non-uni_90_o; (d) non-uni_0_o. i) configuration diagram (FEM), ii) finite element simulation at 1.9 MHz of a simulated 1D wrinkle defect scan on [90/0] CFRP structure for each sensor configuration.

## 4. Experimental study

### 4.1. Materials

A cross-ply CFRP plate of 250 × 25 × 4 mm was fabricated using 32 plies of prepreg (IM7/8552). The layup was $[(90_2/0_2)_4]_s$ and some of the 90° plies had strips cut out, and these strips were then stacked between the first two 90° plies to create an excess of material over which the 0° plies would then appear wrinkled. The plates were cured in the autoclave using the standard cycle and with the aid of a silicone top plate. For more on the complex laminates seeded with asymmetrical out-of-plane waviness please see the work by Maes et al. [42]. With reference to Fig. 6(a), samples G1-3 and G6-G8 all have wrinkles with increasing amplitude but with asymmetry level around 0.2 mm, however G4-5 and G9 specimens have a significant non-zero value of wrinkle asymmetry offset. Fig. 6(b) shows sample G4 as measured using the VHX-7000 N digital microscope (KEYENCE Ltd., UK), where the offset red vertical line highlights the maximum wrinkle amplitude, A, with an 0.6 mm offset from the centre-line (dashed blue line) of the defective zone wavelength ($\lambda$). The amplitudes (wrinkle heights) and skewness measured by micrograph are shown in Fig. 6(c). The wrinkle parameters for each group (G1-G9) of specimens are given in Table 4.

### 4.2. Experimental setup

A schematic diagram of the experimental setup is shown in Fig. 7(a). A Handyscope HS5 (TiePie Engineering, Netherlands) generates a harmonic sinusoidal excitation signal of frequency $f$, delivered to a Howland Current Source (Sonemat Ltd., UK). The current source delivers a consistent current of frequency $f$, to the driver coil to produce a stable magnetic field at each specific frequency measured. The voltage across the driver coil, $V_{mon}$, is used as a reference signal for the received differential pickup voltage of the probe. The pickup coils are connected to a bespoke differential amplifier (see Fig. 7(a)) placed inside a 3D-printed box that houses the entire probe as shown in Fig. 7(b). A direct current (DC) power supply provides a ±5.5 V supply to the differential amplifier. A laptop PC and MATLAB script controls and post-processes the measurement in real-time via serial communication with the Handyscope.

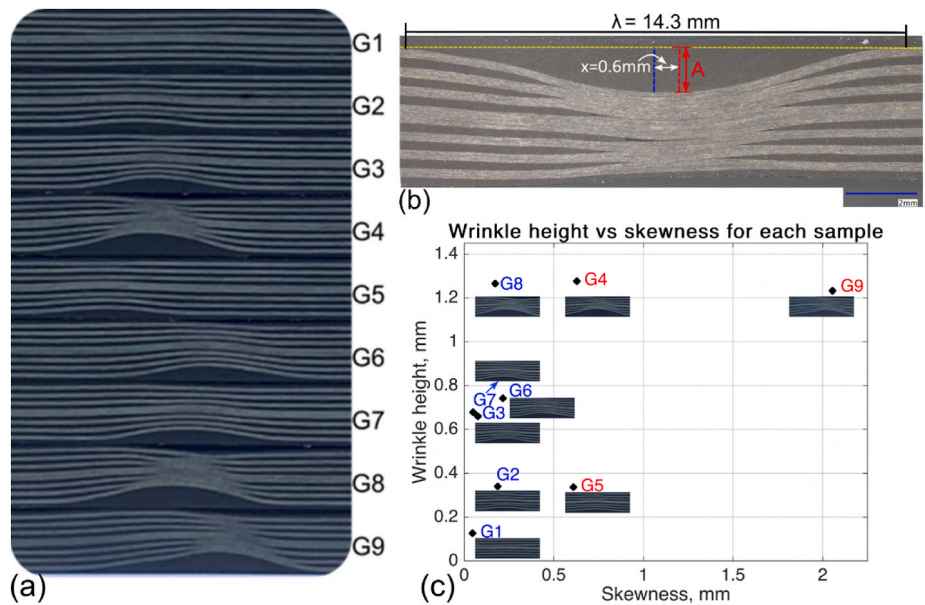

**Fig. 6.** Wrinkle amplitude and skewness for: (a) actual side-view photos of the G1-G9 samples; (b) an example diagram of G7 sample with its wrinkle parameters; (c) plot of each group of specimens in wrinkle amplitude and skewness graph.

**Table 4**
Micrograph measured values of wrinkle amplitude and skewness for each group of specimens.

| Sample | Wavelength [mm] | Height/Amplitude [mm] | Peak offset [mm] |
|---|---|---|---|
| G1 | 12.44 | 0.13 | 0.05 |
| G2 | 10.19 | 0.32 | 0.19 |
| G3 | 11.39 | 0.71 | 0.08 |
| G4 | 14.36 | 1.26 | 0.63 |
| G5 | 10.31 | 0.33 | 0.61 |
| G6 | 12.17 | 0.74 | 0.22 |
| G7 | 12.35 | 0.68 | 0.05 |
| G8 | 12.85 | 1.33 | 0.17 |
| G9 | 14.85 | 1.18 | 2.06 |

A X-LSM200B-E03 Motorized linear stage (Zaber Technologies, Canada) was used to control the linear movement of the specimen with the probe mounted in a static position. The linear stage was placed on a L490/M Lab Jack (Thorlabs Inc., USA) to control the vertical displacement of the specimen. The lift-off variation was measured using a standard feeler gauge and the level of parallelism of probe and sample were controlled by ruler. A transformer-coupled differential amplifier (shown in Fig. 8) was developed for this study, to measure the differential signal between pickup-coils in the sensors. The input transformer coupled amplifier allows a differential measurement of the two receiver coil voltages (i.e. $Vo = Gain \times (V1 - V2)$, where the amplifier voltage gain is equal to 10) to be accurately made by wiring the two probe coils in series opposition and connecting them across the non-ground referenced primary of the transformer. This arrangement provides the required balanced loading effect on the two probe coils as is shown in Fig. 8.

One of the main challenges in establishing any EC inspection technique set-up is minimizing the noise level. There are many sources of noise which limit the detectability of defects in ECT NDE. As well as random electrical noise, coherent noise is also present from sources that include variations in electromagnetic properties of the sample, temperature changes, probe lift-off and tilt of the sensor. As defined in equation (2), the signal-to-noise ratio (SNR) used in this study is defined as the measured peak-to-peak signal of a defective region divided by 2 times the root-mean-squared (RMS) value of the coherent background noise (see Fig. 9), and is used as a metric to quantitatively evaluate probe performance.

### 4.3. Sensor signals

Each scan started by moving the stage with a fixed sample on it along the x-direction while each probe was recording sensor measurements from the static position. The x-axis linear stage with a fixed sample (see Fig. 7(b)) is moved by increments of 0.2 mm. Real-time monitoring of the sensor response was performed via MATLAB with a data saving option for further post-processing. The obtained measurement data is used to calculate the SNR by applying detrending to remove unwanted linear measurement trends resulting from minor lift-off variations. An example of a position-based measurement is illustrated in Fig. 9. The first 0.5 mm scan is performed while the sensor was moving on the sample by increments of 2.5 $\mu$m in the non-defective region and is used to assess the random electrical noise in the measurement. The measurement range between 80 and 119 mm shows the wrinkle defect signal response (see the vertical blue dashed line in Fig. 9(b)) while the rest of the scan corresponds to the coherent structural information about the sample (referred to as structural noise). The horizontal red dashed lines indicate the ±RMS noise bounds (see Fig. 9(b)) of the structural noise. The probe is sensitive to both the structural material background noise and the defective region. Fig. 9(c) shows an example plot of G8 sample's data using four probe configurations.

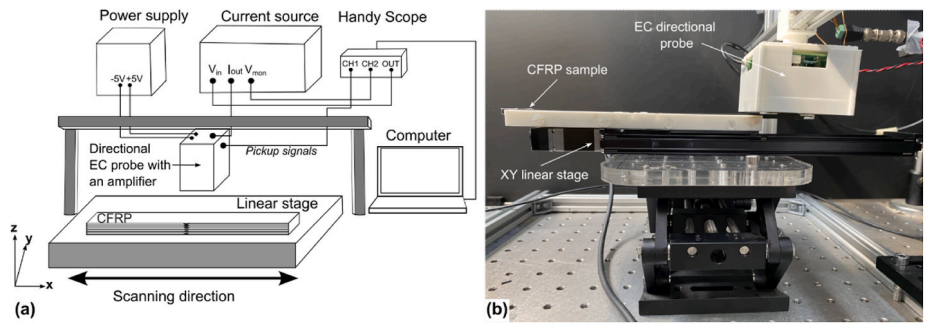

**Fig. 7.** (a) Schematic illustration of experimental setup and (b) actual photo of the experimental scanning system (side view).

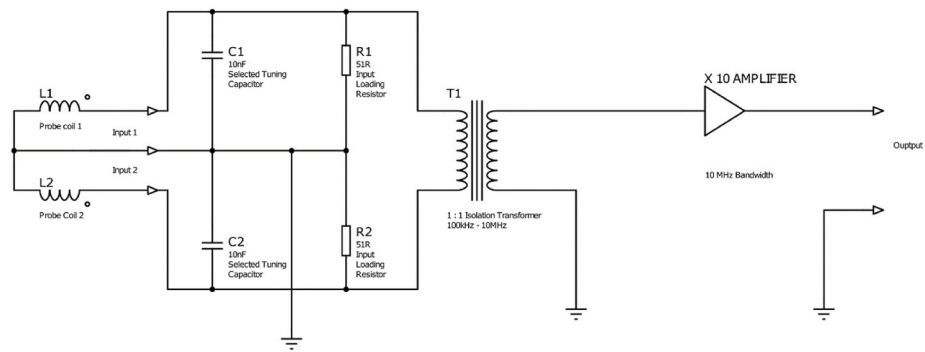

**Fig. 8.** The simplified block diagram of the transformer coupled differential amplifier.

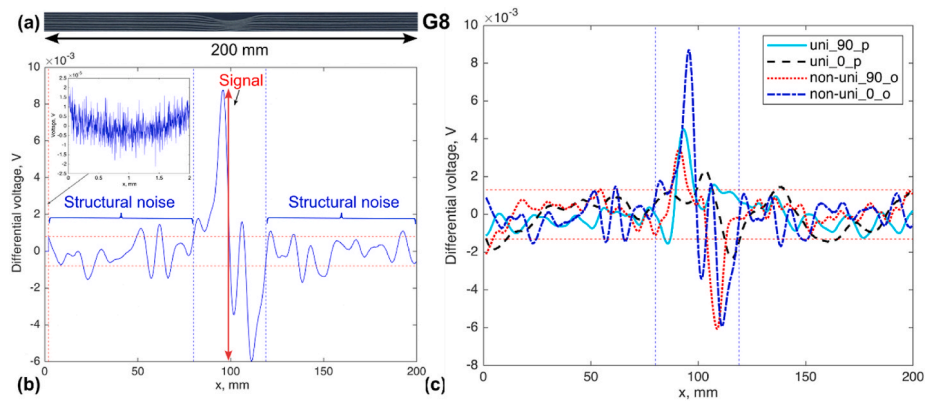

**Fig. 9.** (a). Side view image of CFRP sample, example position-based plot of sample G8 measured at 1.6 MHz using (b) the non-uni_0_o probe showing (in set) sensor and structural noise, and c) four sensor configurations.

The aim of the paper was to demonstrate the validity of a simplified finite element model in comparison to experimental results. The results generally agree across three primary qualitative considerations: 1) In both simulated and experimental results, the periodic (background coherent structural noise) fluctuation is discernible, a critical factor when evaluating the signal to noise ratio. 2) When the probe scans over the wrinkle, positive and negative peaks are observed in the signals, for both the simulated and experimental datasets. 3) The G8 sample from the experimental study results is the most severe defect and could be equated to the 75 % drop in conductivity within the virtual defect, both showing the largest differential response. Despite these similarities, further refinement of the modelling approach is required, and will be the subject of future studies under development. This could be achieved by providing a more realistic simulation of electrical and dielectric variations for high-frequency eddy current studies, and also by calibrating out circuit components representing parasitic components present in our experimental circuit. Upon a close comparison of the data in Figs. 9 and 5 ii, the following observations were made: Both sets of data exhibit sinusoidal fluctuations, identifiable as background coherent structural noise. This observation is vital for evaluating the signal-to-noise ratio, as it allows for a better understanding of underlying patterns in the data. During the scanning of the probe over the wrinkle, positive and negative peaks are noticed within the signals. These peaks occur consistently in both simulated and experimental datasets, indicating uniform behavior across different conditions. The G8 sample from the experimental study represents the most severe defect. It could be equated with a 75 % drop in conductivity in the virtual defect. Both scenarios demonstrate the largest differential response, highlighting the crucial aspect of defect detection. The amplitude of signals across both simulation and experimental data falls within the range of $10^{-2}$ [V] to $10^{-3}$ [V], signifying a consistent behavior between the two datasets. However, it is essential to note the differences between the modeling and experimental data: For instance, non-uni_0_o has the largest G8 response in the experimental data, while non-uni_90_o exhibits the most significant response in the simulation data. The latter seems more logically correct, as non-uni_90_o induces currents in the fiber direction. Besides, the modeling also shows an asymmetry in Uni_90_p, with signal amplitudes reduced to $10^{-5}$ [V]. This could be attributed to the dominance of noise generated by the minor asymmetry of the mesh in this particular case. Therefore, there seems to be a possibility of missing certain physics phenomena in the modeling. This shortfall can be addressed by implementing a more realistic simulation of electrical and dielectric variations tailored for high-frequency eddy current studies, utilizing a strictly controlled 3D symmetric mesh, and integrating with circuit components to emulate the parasitic components present in our experimental circuit. In conclusion, while the comparison between Figs. 9 and 5 ii has led to valuable insights, it also emphasizes the necessity of further refinements in the modeling process to bridge the gap between the simulated and experimental data.

### 4.4. Results & discussion

A series of studies were conducted to evaluate the sensitivity of Tx-dRx probe configurations and explore the detectability of out-of-plane wrinkling. An example illustration of the simple data processing method used is presented in Appendix A1. The first experimental study describes the evaluation of excitation frequency on measurement sensitivity for the 4 probe configurations (see Fig. 2), by examining the SNR and measurement uncertainty when measuring wrinkle defect.

Experimental validation was performed to comprehensively evaluate the probe configuration capabilities. Kosukegava et al. [26] highlighted the potential for sensitivity improvements in their Tx-dRx probes when operating beyond the resonance of the pickup coil. The experimental study took five repeat scans of sample G9 for each frequency between 1.4 and 2.3 MHz, with increments of 0.1 MHz and a 0.2V peak-to-peak excitation voltage. Consequently, the next study shows an evaluation of the detectability of out-of-plane wrinkles for each probe configuration at different frequencies around resonance. The final study shows the relationship between wrinkle height and SNR in the probe configurations.

The experimental results of the frequency evaluation study are plotted in Fig. 10, showing $V_{pp}$ and $N_{RMS}$ as a function of frequency, along with the random measurement noise contribution. These are compared to the SNR (see equation (2)) for a single out-of-plane wrinkle height (sample G9). The highest SNR was found to exist just beyond the resonance of each probe. To evaluate the reason for this peak, the defective signal, probe and coherent noise are shown separately as a function of frequency (the green vertical line represents resonant frequency of each probe) in Fig. 10 i. This demonstrates that for each configuration, the coherent noise and defect signal follow the same trends but a slightly shifted in the frequency domain, such that larger differences in their voltages occur. The coherent background noise was the most significant factor influencing to SNR level at each frequency. The most sensitive probe to coherent noise was found to be uni_90_p, while the most sensitive to defects was non-uni_90_o probe. The salient metrics from the raw EC data (please see Appendix B1) have been extracted and plotted in Fig. 10 ii. The diagrams in Fig. 10 ii show that the probes with the driver coil parallel (p) to the surface of the sample (i. e. generating a uniform magnetic field beneath the pickup coils) achieves lower SNR results than the orthogonal (o) configurations as predicted in the FEM (Fig. 10 iii).

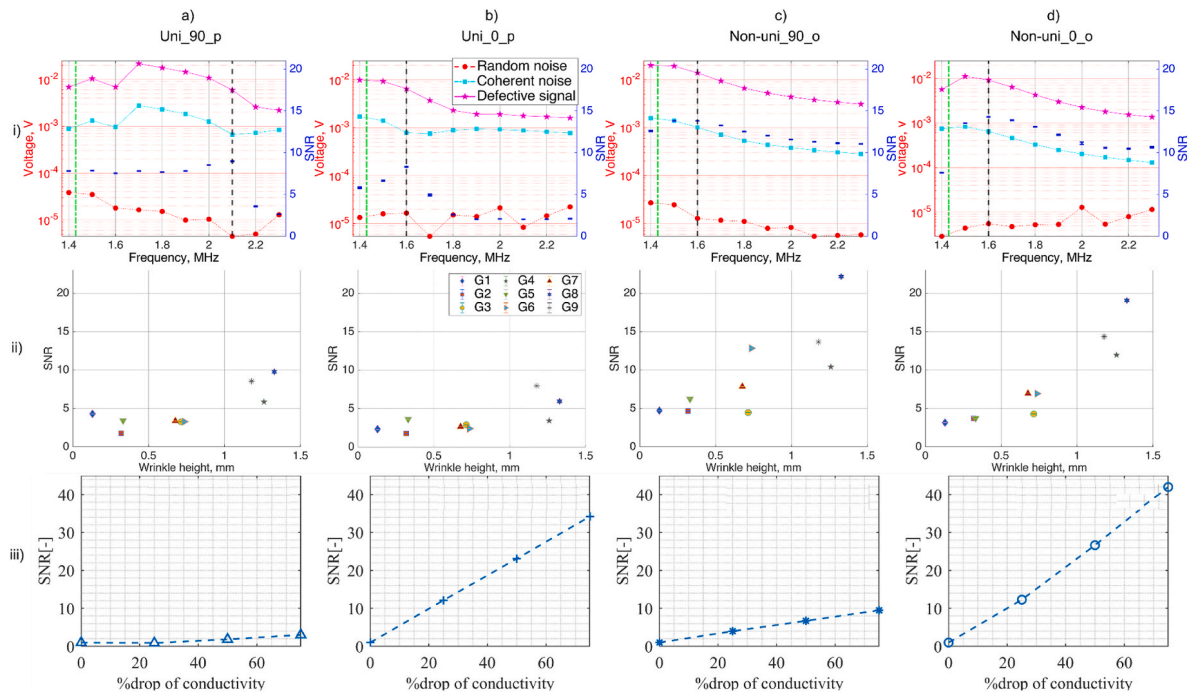

**Fig. 10.** Signal-to-noise results for sensor configurations (a) uni_90_p; (b) uni_0_p; (c) non-uni_90_o; (d) non-uni_0_o. i) Experimental study of response as a function of frequency showing defect signal compared to coherent and measurement (random) noise, and signal-to-noise ratio (blue), ii) signal-to-noise at optimum sensor frequency as a function of wrinkle height, and iii) finite element modelled SNR for changes in conductivity to simulate wrinkles. (For interpretation of the references to colour in this figure legend, the reader is referred to the Web version of this article.)

The sensitivity to skewness, which has previously been shown by Maes et al. [42] to be relevant to the local laminate strength, can be assessed by evaluating wrinkles with comparable height but differing skewness (see Table 4) and by comparing them in terms of their SNR change (see Fig. 10 ii). G3/G6 and G2/G5 samples have comparable wrinkle heights but have triple and fivefold increases in terms of skewness value respectively. Non-uni_0_o shows that the SNR for these wrinkles are comparable, while non-uni_90_o probe results show separation in the SNR with increased skewness. However, it is not possible to distinguish between high skewness and low wrinkle height. Based on the experimental results, the most suitable probe for out-of-plane wrinkling detection was the non-uni_0_o, showing a slightly better performance in targeted defect detection in comparison with non-uni_90_o, and insensitivity to skewness. For the simulation study, the best configuration for detecting out-of-plane wrinkles is also non-uni_0_o, demonstrating better linearity compared with non-uni_90_o despite the fact that it has a higher SNR value for G9. The uni_0_p and non-uni_90_o probes do not demonstrate the expected performance in the simulated study, which is most likely due to the simplicity and assumptions of the modelling approach. Good agreement however, is observed between FEM and experimental study SNR results presented for out-of-plane wrinkles in probes uni_0_p and non-uni_0_o.

## 5. Conclusion

An eddy-current (EC) directional probe with asymmetric transmit and differential receive (Tx-dRx) coils was designed, constructed and characterized to evaluate the detectability of out-of-plane wrinkling in carbon fibre composite structures. The results of an experimental study demonstrate improved signal-to-coherent-noise (SNR) is achievable through careful selection of the asymmetric driver coil's relative orientation. The experimental results were compared to EC Tx-dRx finite element modelling, applying a new spatially-modulated conductivity tensor approach to simulate tow/ply structural noise, showing good agreement in trends observed between different sensor configurations. As such future work will be able to employ less experimental testing and can implement finite element model optimization of sensor designs.

High-sensitivity to wrinkle-height was achieved when the driver-coils B-field was orthogonal to the sample surface. In this way, we have demonstrated a potential method for designing real-time ECT monitoring sensors for wrinkle detection in CFRP manufacturing with high SNRs which could have a significant impact on manufacturing rates in smart manufacturing for industry 4.0.

**Authorship contributions**

Meirbek Mussatayev: Writing - Original Draft, Conceptualization, investigation, data analysis, visualization, methodology, project administration, and Writing - Review & Editing; Qiuji Yi: Conceptualization, Finite Element Modelling and its draft writing; F, Mark Fitzgerald: amplifier design; Vincent K. Maes: sample preparation; Paul Wilcox: Conceptualization, manuscript revision, supervision; Robert Hughes: Project administration, methodology, conceptualization, manuscript revision, supervision.

**Declaration of competing interest**

The authors declare that they have no known competing financial interests or personal relationships that could have appeared to influence the work reported in this paper.

**Data availability**

Data will be made available on request.

**Acknowledgements**

MM was funded by a Bolashak International Scholarship from the Kazakstani Government. This work received funding from EPSRC project CerTest - Certification for Design - Reshaping the Testing Pyramid (EP/S017038/1) and EPSRC Future Composites Manufacturing Research Hub (Grant: EP/P006701/1). The data created during this project can be found at the University of Bristol Research Data Storage Facility (RDSF) - DOI: 10.5523/bris.nsyfe0o66hs61znpr5ncp08n8.

**Appendix A. Supplementary data**

Supplementary data to this article can be found online at https://doi.org/10.1016/j.compositesb.2023.111048.

**Appendix A1**

*Signal processing of in-line EC data*

The simple signal processing of in-line EC data is applied for the given study. The robust method of distinguishing the out-of-plane wrinkle signal from both coherent background noise of the CFRP material and the probe configuration is an important prerequisite to calculate the SNR properly. The data was post-processed to detrend the obtained signal amplitude, the scan positions of the defective signal and coherent background noise zones were identified, and finally the SNR ratio was calculated. The post-processing method can be further justified as a step-by-step algorithm:

- Baseline subtraction
- Detrend
- Peak detection

The experimental EC data has offset from zero. Thus, it was removed by identifying the mean of the experimental data then subtracting the mean value from the experimental. To eliminate any unwanted background trends a low-order polynomial is fitted and subtracted from the signal via polyfit and polyval MATLAB functions respectively. After detrending the signal MATLAB's findpeaks function was applied to find the local peak's width automatically. The peak detection algorithm found the width of peaks that have 3 x RMS whole signal positions. Finally, by multiplying the width of the local maxima of the signal two times the signal positions were defined. The rest of the signal trends were assigned as the structural background noise of the non-defective zone.

The evaluation of sensitivity to the wrinkle area was calculated from the signal to coherent noise. The peak-to-peak value of the signal positions divided by the 2 x RMS value of the coherent noise structural noise.

Fig. 9 c shows the example plot results of G8 sample using signal processing of in-line EC data. The signal starts and ends indexes (see the vertical blue dot lines in Fig. 9c) were identified by finding the cross section between the mean RMS values of coherent structural noise and the width of the local peak. The maximum and minimum noise threshold were defined by two time multiplication of the mean of RMS noise value at each frequency for selected sample (see parallel blue dote lines in Fig. 9c).

**Appendix B1**

Raw EC data for wrinkle height assessment

running

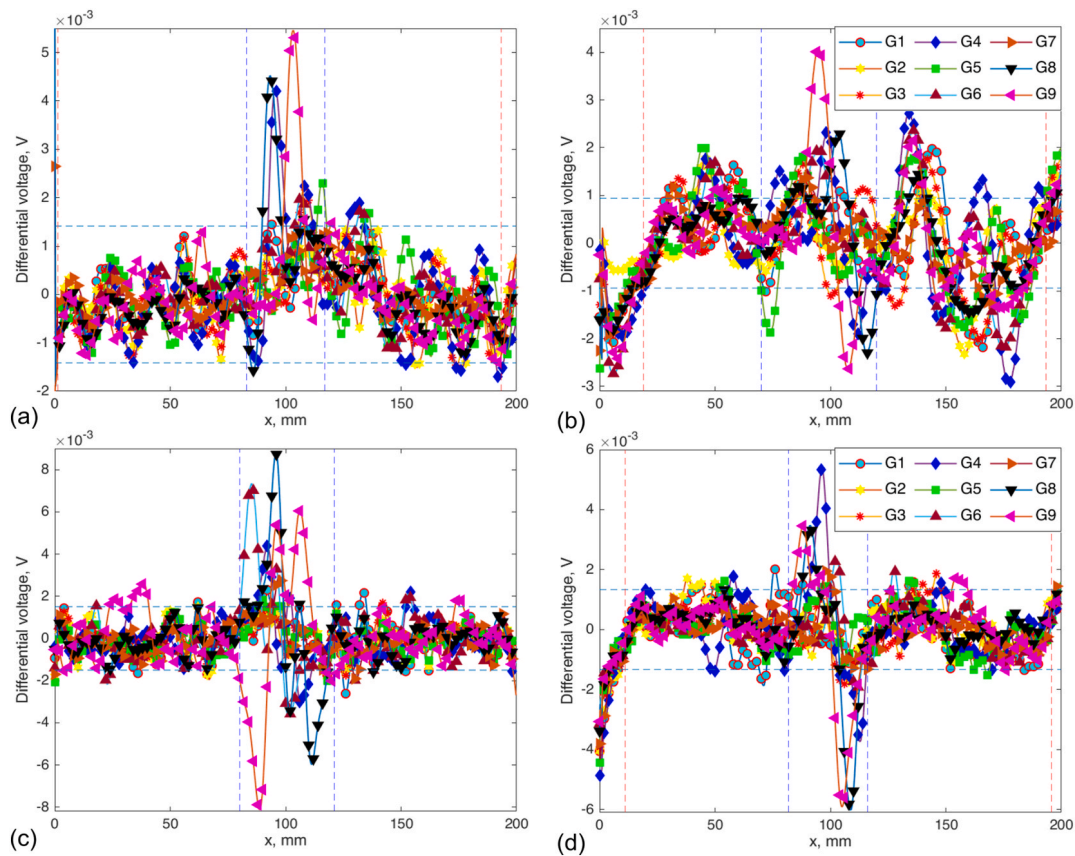

**B1.Fig. 1.** he raw EC scan data for wrinkle height assessment using four sensor configurations: (a) uni_90_p; (b) uni_0_p; (c) non-uni_90_o; (d) non-uni_0_o.